# Multi-Objective Recommendations: A Tutorial

Yong Zheng[1], David (Xuejun) Wang[2]
[1]Illinois Institute of Technology, Chicago, IL 60616, USA
[2]*Morningstar Inc., Chicago, IL 60602, USA*

***Abstract:*** Recommender systems (RecSys) have been well developed to assist user decision making. Traditional RecSys usually optimize a single objective (e.g., rating prediction errors or ranking quality) in the model. There is an emerging demand in multi-objective optimization recently in RecSys, especially in the area of multi-stakeholder and multi-task recommender systems. This article provides an overview of multi-objective recommendations, followed by the discussions with case studies. The document is considered as a supplementary material for our tutorial on multi-objective recommendations at ACM SIGKDD 2021.

***Keywords:*** recommender system, multi-objective, multi-objective optimization

# Table of Contents

# I. Introduction

This article is utilized as a supplementary material for our tutorial on multi-objective recommendations[1] at ACM SIGKDD 2021. There are two major sections – Section II provides an overview of the multi-objective optimization (MOO) methods, including the definition of multi-objective problems and the solutions, the notion of dominance relation, mainstream multi-objective optimization technologies, as well as the corresponding challenges, such as how to select a single optimal solution from the Pareto set. Section III covers the introduction and discussion on multi-objective recommendations. More specifically, we identify six contexts in which we may need MOO in the area of recommender systems, and discuss each of them by revisiting the underlying motivations, objective definitions and case studies.

Therefore, this document can be considered as the extended version of our KDD tutorial. We also published a survey paper, see below.
- Zheng, Y., & Wang, D. (2021, August). "*Multi-objective recommendations*". In Proceedings of the 27th ACM SIGKDD Conference on Knowledge Discovery & Data Mining (pp. 4098-4099).
- Zheng, Y., & Wang, D. X. (2022). "*A survey of recommender systems with multi-objective optimization*". Neurocomputing, 474, 141-153.

# II. Multi-Objective Optimization

In this section, we deliver the overview of the MOO methods

## 1. Background and History of MOO

Many people may not realize that Multi Objective Problems (MOP) could be general in our life and society. Every day we are making decisions trying to compromise among different objectives. If you were a car buyer, you want to lower price and gas consumption, and higher level of comfort and performance. With limited budget, these objectives are conflict each other. You need to balance between lower price or gas consumption and comfort or performance. On the society level, the central bank's monetary policy needs to balance among inflation rate, unemployment rate, trade deficit and other economic factors. Lower interest rate may reduce the unemployment but may increase the risk of inflation at same time. We can find similar situations in other areas such as engineering and communication designs. In each situation, the decision maker (DM) wants to optimize more than one objective, which are conflict each other in most of cases.

1.1 MOO examples in Finance

If an investor has certain amount of annual income, how does this investor invest money in child education, house buying and retirement saving, as well as improve daily life quality. This is a traditional *Financial Planning Problem*. The investor here is a Decision Maker (DM). If he/she spends too much money in travel or dinning out to maximize the enjoyment of daily life, there

---
[1] https://moorecsys.github.io/

would be less money to invest in education, house buying and retirement saving. In this investor's mind, all objectives should be maximized and again all these objectives are conflict each other, meaning improvement of one objective may downgrade other objectives. In this prospect, a financial planning problem is a multi-objective optimization problem. Figure 1 is an illustration of this problem:

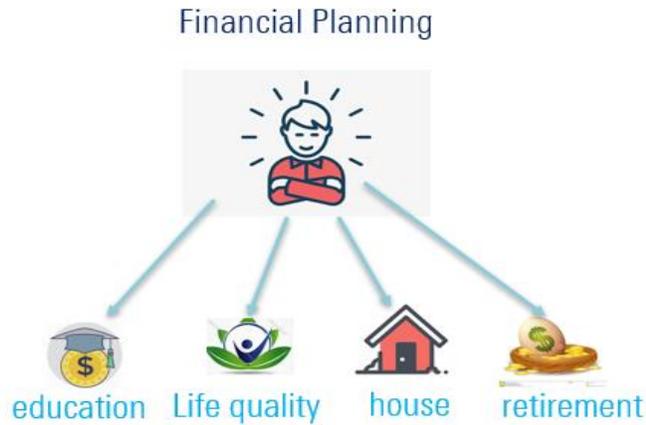

Figure 1. Example of Multi-Objective Problem in Financial Planning

Another well-known multi objective problem in finance field is the *Portfolio Optimization* problem. A portfolio is a combination of different stocks, mutual funds other securities available in financial markets. In this problem, an investor needs to construct a portfolio that minimizes risk and maximizes return. However, it is well known that a lower risk portfolio will has lower expected return. Portfolio optimization process can generate a series of efficient portfolios that has different level of risk and return. The image of these efficient solution in Risk/Return plan is called *efficient frontier*, as illustrated in Figure 2:

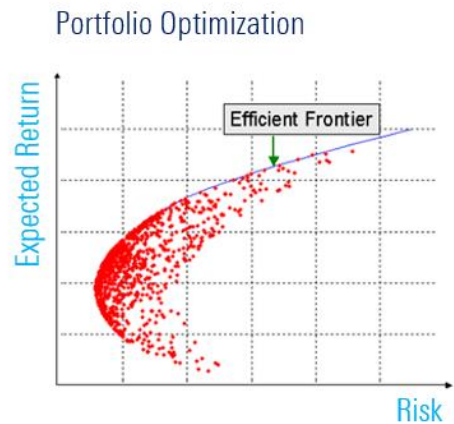

Figure 2. Efficient Fronter of Portfolio Optimization

The investor can choose a portfolio along the efficient fronter line based on his/her preference on level of risk and return.

1.2  MOO examples in Recommender Systems

In traditional recommender systems, the goal is to generate a list of items that most likely meets the user's need. There are many different metrics to evaluate user's need. Here are three most used metrics during the offline evaluations:

- *Accuracy*: measure the difference between recommended items and the expected items
- *Diversity*: measurement of un-similarity between recommended items. Maintaining a certain level of diversity increase the chance for user to choose useful items from the list
- *Novelty*: measure the likelihood that a recommender system to generate recommendations that user may not be aware of.

When a recommender system tries to maximize all three metrics at the same time, it become a multi objective optimization problem. Each metric is an objective. In this case, increasing accuracy may reduce the diversity and novelty, since more items that user assessed in the testing data set are in recommendation list. Increasing diversity or novelty may also reduce the accuracy for the same reason. We have a recommendation system that tries to optimize multiple objectives that conflict each other.

If a recommender system that optimizes multiple metrics is an artificial situation modeled by researchers, the multi-stakeholder recommendation is a real-world problem with multiple objectives. Traditional recommendation system only considers the interest of end users. A multi-stakeholder recommender system needs to generate recommendation for end users as well as other stakeholders. Such examples can be found in commercial online market platforms, such as Amazon.com, Wallmart.com. The sellers and platform owners are two stakeholders along with end users.
In multi stakeholder recommender systems, the utilities of all stakeholders (end users, item sellers and platform owners) need to be maximized at same time. The challenge here is that increasing utility of one stakeholder may reduce the utilities of other stakeholders. For example, if the recommendation includes more expansive items, the seller's utilities (profit) may increase at the cost of user's utilities (items likely to buy). In conclusion, a multi-stakeholder recommender must deal with multi objectives that conflicts each other.

1.3  MOO history and trend

Francis Ysidro Edgeworth, an Irish philosopher and political economist in late 1800's, published a very influential book, *Mathematical Psychics: An Essay on the Application of Mathematics to the Moral Sciences*, in 1881. In his book, Edgeworth studied the multi-utility problem with two hypothetical consumer criteria, $P$ and $\Pi$: "*It is required to find a point $(x, y)$ such that, in whatever direction we take an infinitely small step, $P$ and $\Pi$, do not increase together, but that, while one increases, the other decreases*".  Almost 20 years later in 1906, a French economist, Vilfredo Pareto, studied multi objectives optimization problem in a more formal way in his book, *Manual of Political Economy*, in which the concept of Pareto Optimal was formally defined.

Although the concept of Pareto optimal in economics was introduced in early 1900, the Multi Objective Optimization (MOO) only became a dedicated research area after 1970's. MOO study enjoyed a temporary flourish in mid-1980's and gained a dramatic development after year 2000 and thereafter. Some of those underlying driving factors are related to new computing technologies and wide applications of machine learning and artificial intelligence.

## 2. Concepts and Definitions of Multi Objective Optimization

To formally introduce basic concepts of MOO, we need to define the MOO problem using a set of objective functions and constraints:

$$\min_{x}(f_1(x), f_2(x), \ldots, f_M(x)) \tag{1}$$

Subject to
$$g_j(x) \geq 0, j = 1, 2, \ldots, J$$
$$h_k(x) = 0, k = 1, 2, \ldots, K$$
$$x_i^L \leq x_i \leq x_i^U, i = 1, 2, \ldots, n$$

Here $x \in R^n$ is a $n$-dimension decision variable. $f_i(x)$ is $i$th objective function. $M$ is the number of objectives. The other three equations or inequalities in the above define the feasible value that $x$ can take. In the following discussion, we use $S$ to denote the set of all *feasible solutions*. Then we have:

$$S = \{x \mid x_i^L \leq x_i \leq x_i^U, \ g_j(x) \geq 0, h_k(x) = 0, j = 1, 2, \ldots, J, k = 1, 2, \ldots, K, i = 1, \ldots, n\}$$

We also define a vector function $F(x) = (f_1(x), f_2(x), \ldots, f_M(x))$ in objective space. With this notation, a MOO problem (1) can be defined as:

$$\min_{x \in S} F(x) \tag{2}$$

*Example 1.* A recommender system that maximizes both accuracy and diversity:

- Decision variable $x$: a top $N$ recommender list
- Feasible solution set $S$: all possible top $N$ recommender lists
- First objective $f_1(x)$: 1 – accuracy
- Send objective $f_2(x)$: 1 – diversity
- Vector objective $F(x) = (f_1(x), f_2(x))$

Then the recommendation list that maximize both accuracy and diversity can be expressed by

$$\min_{x \in S} F(x) \quad \text{or} \quad \min_{x \in S}(f_1(x), f_2(x)) \tag{3}$$

As we discussed in previous section, a special challenge of finding a good solution with multi objectives is that these objectives are conflict each other. Because of the conflict nature of these objectives, it is not easy to determine which solution is better. We use next toy example to show this challenge:

*Example 2.* Let
$$f_1(x) = 2(x-1) + 1$$
$$f_2(x) = 2(x-3)^2 + 1$$

Find solutions in $[0, 6]$ that minimizes both $f_1$ and $f_2$:

$$\min_{x\in[0,6]}(f_1,f_2)$$

We plot both $f_1$ and $f_2$ in the following Figure 3:

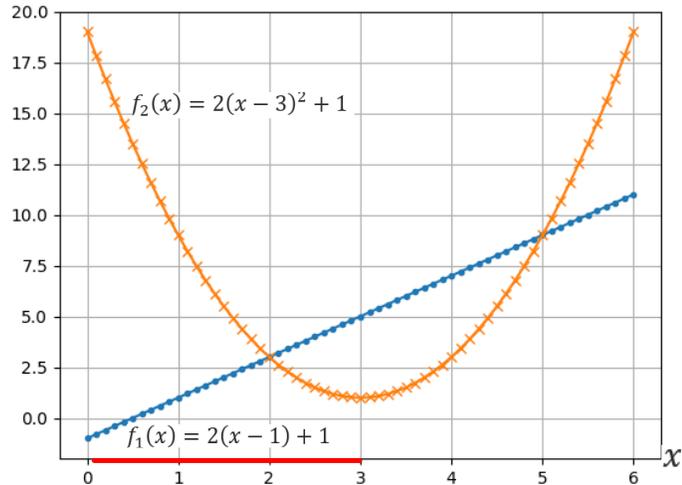

Figure 3. Two Conflicting Objectives

In interval of $[3, 6]$, both $f_1$ and $f_2$ are increasing. Obviously, $x = 3$ is the best solution in $[3, 6]$. However, in interval of $[0, 3]$, $f_1$ is creasing, but $f_2$ is decreasing. By changing $x$ in $[0, 3]$, we cannot determine which solution is better, since a decreasing one objective will make the other objective bigger. A typical multi objective dilemma!

2.1 Definitions:

To solve the above dilemma, we need to introduce a new concept to rank the solutions with multi objectives.

**Definition**: (*Dominance Relation*) A feasible solution $x \in S$ of problem (2) is said to be *dominated by* another solution $x^*$ if and only if

$$f_m(x^*) \leq f_m(x) \text{ for all } m = 1,2,\dots,M$$

and there exists at least one $m'$ such that:

$$f_{m'}(x^*) < f_{m'}(x)$$

The first set of inequalities makes sure all objectives of $x^*$ are not bigger than those of $x$. The second inequality guarantees that they do not have same objective values. In Figure 4, we can see that both solution $A$ and $B$ dominate $C$ and $D$. But there is no dominance relationship between $A$ and $B$, as well as between $D$ and $B$.

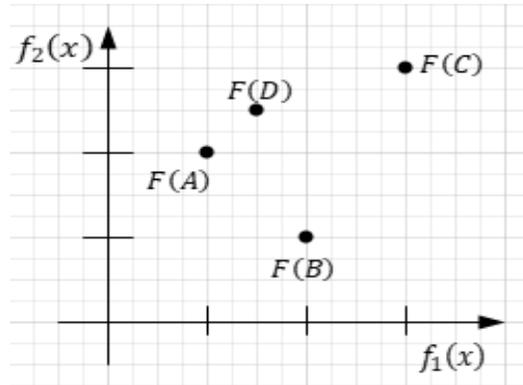
Figure 4. Dominance Relations

With the definition of dominance relationship, we introduce the most important concepts in MOO:

**Definition**: (*Pareto Optimal, Non-Dominated*) A solution that is not dominated by other solutions are called *non-dominated solution* or *Pareto Optimal Solution.*

**Definition**: (*Pareto Set*) A set of all Pareto optimal solutions. We use $P$ to denote Pareto set in this paper if not explained.

In the following discussion, Pareto optimal will be used interchangeably with Non-dominated solution.

**Definition**: (*Pareto Front*) The image of objective vector function $F(x)$ from all solutions in Pareto set is called *Pareto Front* in objective space

In the above example of the 2-metric recommender system (*Example 1*), we can identify Pareto set and Pareto front in Figure 5:

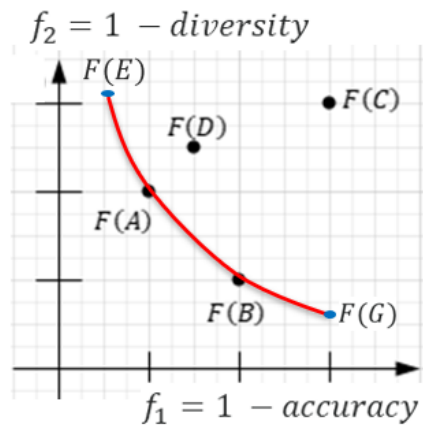
Figure 5. Pareto Front of Example 1

In Figure 5, solution $A, B, G, E$ are Pareto optimal. Any solution $x$ that maps the objective vector $F(x)$ onto the curve connecting $F(E), F(A), F(B), F(G)$ is Pareto optimal. All such Pareto optimal solutions constitute the Pareto set. The red curve from $F(E)$ to $F(G)$ is the Pareto front.

For the second toy example above (*Example 2*), we plotted all the points in objective space for feasible solutions $x \in [0, 6]$, as shown in the Figure 6:

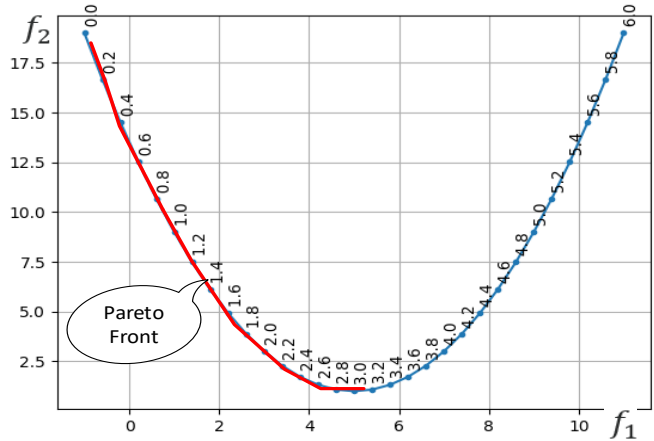

Figure 6. Pareto Front of Example 2

In Figure 6, the corresponding solution value $x$ is marked on the curve. It can be observed that all solutions in $[3, 6]$ is dominated by at least one solution in $[0, 3]$. No solutions can dominate solutions in $[0, 3]$. Then $[0, 3]$ is the Pareto set. The corresponding curve (in red color) is the Pareto front.

With these newly introduced concepts, we can restate the MOO problem:

*For a MOO problem defined in (2), we need to find a single Pareto optimal solution or all the Pareto Optimal solutions (the Pareto set)*

2.2 MOO decision making process

A question will be asked following the above discussion: under which situation we need to find one Pareto optimal solution, and under which situation we need to find the whole Pareto set? The answer relies on how the Decision Maker (DM) are involved in the multi objective problem decision process. There are four types of DM involvement in the MOO decision process:

- *A Priori:* DM sets up preference weight for each objective
- *A Posteriori*: DM chooses the best solution from a set of Pareto optimal solutions
- *Interactive*: DM interacts with MOO algorithm during each step to choose the best solution
- *No DM*: DM is not available in the whole MOO process

In a priori case, since DM can set up preference of each objective, the MOO problem can be easily transformed into a new single objective by summing up all objective with preference as weights. There is only one best solution after finding the solution the minimize the newly created single objective. However, in a posteriori case, DM cannot set objective preference at the beginning. Instead, DM will choose one 'best' solution from all possible best solution, which is all non-dominated solutions from MOO algorithm. In this case, a Pareto set must be produced from MOO algorithm. In the third case above, DM interacts with MOO algorithms in some middle steps, so DM can make decision during the MOO process. The result is a single Pareto optimal solution. When DM is not available during the whole process, not only the MOO algorithm needs to produce the Pareto set, a 'best' solution must be selected from the Pareto set for DM.

To summarize, both a priori and interactive case will produce one Pareto optimal solution for DM. A posteriori case will produce a Pareto set, so DM can choose the one he/she likes. If no DM is available, MOO algorithm needs to generate Pareto set and find a 'best' solution for DM.

2.3 Categories of MOO algorithms

All MOO algorithms can be classified into two categories:

- Scalarization algorithms
- Multi Objective Evolutionary Algorithms (MOEA)

The central idea of scalarization algorithms is to transform multi objectives into one objective. If DM preference is available, this transformation can be done easily, and a single solution is produced. If DM preference is not available, such as in a posteriori and no DM cases, multiple runs of scalarization algorithm with different parameters are needed to find the Pareto set.

MOEA follows the natural evolution process, such as gene evolution, a flock of birds find best path seeking food and other resources, a cooling process of melted crystal, and many others. All EA methods are heuristic and have minimal requirements for the objective functions. Since all EA methods are population based, it can generate a Pareto set in one run under proper settings. Because of this advantage, MOEA is widely used in a posteriori and no DM cases.

# 3. Scalarization Algorithms that Solve MOO

There are many different types of scalarization algorithms that solve the MOO problem. The difference between them lies on how to transform the multi objectives into a single objective. After transforming multi objectives into one objective, the problem can be solved using regular optimization algorithms. Here is a list of commonly used methods of scalarization algorithms:

- Weighting methods
- $\epsilon$-constraint method
- Normal Boundary Intersection (NBI) & Normal Constraint (NC) method
- Goal Programming
- Physical Programming
- Lexicographic Method

In this section, we will explain first three approaches in details: Weighting methods, $\epsilon$-constraint method and Normal Boundary Intersection (NBI) & Normal Constraint (NC). These three methods are simple to implement and more popular than others. Interested readers can check related references for other methods.

3.1 Weighing methods

Among various form of weighting methods, the *Weighed Sum Method* is the most simple and popular scalarization method. Given a set of weights, $w_i$, for each objective, the multi objectives are combined into one objective by a weighted sum:

$$\min_{x \in S} \sum_{i=1}^{M} w_i f_i(x) \qquad (4)$$

where $\sum_{i=1}^{M} w_i = 1$ and $w_i > 0$. The weight condition here guarantees that the solution of (4) is the Pareto optimal solution of the original MOO problem (2) [1]. By setting different weight vector $(w_1, w_2, \ldots, w_M)$ we can get different Pareto optimal solutions. The question is if we can find all Pareto solutions (Pareto set) by selecting different weight vector in (4). this is only true if the original problem (2) meets the following convex conditions [1]:

- The feasible solution set $S$ is convex
- Each objective function $f_i(x)$ is convex

The MOO problem (2) that satisfies these two conditions is said to be *convex.* In theory, if the MOO problem is convex, we can find all Pareto set by changing weights in (4). In practical applications, we can only select limited number of weights and the number of Pareto solutions is also limited. The goal is to find evenly distributed Pareto solutions in objective space. Unfortunately, evenly distributed weights may not produce evenly distributed Pareto solutions [2]. There are different mechanisms to systematically changing the weight to get an evenly distributed Pareto set [1].

If the MOO problem is not convex. There are cases that some Pareto solutions cannot be found no matter which weight is used in (4).

*Example 3:* (Non-Convex MOO) Let

$$f_1(x) = x_1$$
$$f_2(x) = 1 + x_2^2 - x_1 - 0.1\sin(3\pi x_1)$$

where $\leq x_1 \leq 1$ and $-2 \leq x_2 \leq 2$.

The weight change and the corresponding Pareto solutions can be found in the following dynamic graph Figure 7 (https://commons.wikimedia.org/wiki/File:NonConvex.gif):

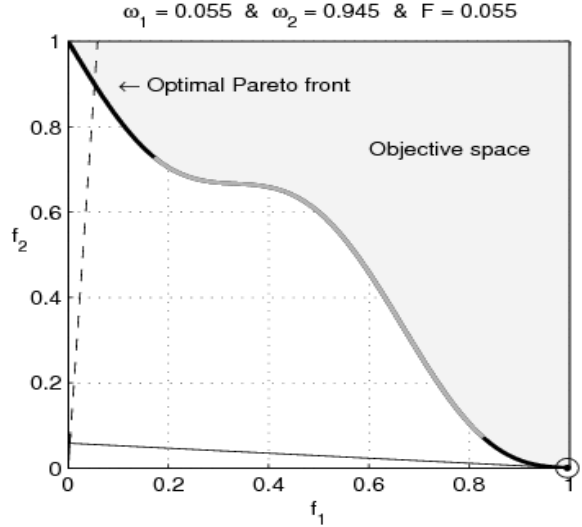

Figure 7. Non-Convex Pareto Front

In this example, the solutions in the concave part of the Pareto front cannot be found under any weight settings.

A variation of weighted sum method is *Weighted Exponential Sum*:

$$\min_{x} \sum_{i=1}^{M} w_i [f_i(x)]^p \tag{5}$$

where $1 \leq p < \infty$, $\sum_{i=1}^{M} w_i = 1$ and $w_i > 0$. Under the weight condition, the solution of (5) is always Pareto optimal. For convex problem, it can find complete Pareto set. If the MOO problem (2) is not convex, increasing $p$ can increase the chance of final all Pareto solutions by changing weights [3].

*Weighted Metric Methods* [4] is a weighted method that compares the distance to the ideal solution in objective space. There are three ways to define an ideal solution:

- Utopia point: a point that takes minimum of each objective
- Goal point: a point consists of ideal values specified by DM for each objective
- Origin point: zero for the ideal point

Let $F^* = (f_1^*, f_2^*, \ldots, f_M^*)$ be the ideal point. $f_i^*$ can take the value using one of the three methods. The weighed metric method can be expressed by:

$$\min_{x} \left[ \sum_{i=1}^{M} w_i^p |f_i(x) - f_i^*|^p \right]^{\frac{1}{p}} \tag{6}$$

where $1 \leq p < \infty$, $\sum_{i=1}^{M} w_i = 1$ and $w_i > 0$. Weighted metric method has the same properties regarding finding the Pareto solutions and Pareto set.

If we take $p \to \infty$ in (6), we will get *Weighted Chebyshev Method* [5]:

$$\min_{\mathbf{x}} \max_{i} \{w_i |f_i(x) - f_i^*|\} \tag{7}$$

where $\sum_{i=1}^{M} w_i = 1$ and $w_i > 0$. The biggest advantage of weighted Chebyshev method is that it can get a complete Pareto set by changing weights without convex requirement [6]. However, it may get non-Pareto solutions in some cases.

The other weighting methods include *Exponential Weighted Criterion* method [7]:

$$\min_{\mathbf{x}} \sum_{i=1}^{K} (e^{pw_i} - 1) e^{pf_i(x)}, p \geq 1. \tag{8}$$

and *Weighted Product Method* [8]:

$$\min_{\mathbf{x}} \prod_{i=1}^{K} |f_i(x)|^{w_i} \tag{9}$$

These methods may encounter computation overflow issues since the objective are involved of exponential or product computations.

To summarize the above weighting methods, here are some major observations:

- The weight condition $\sum_{i=1}^{M} w_i = 1$ and $w_i > 0$ can guarantee to get Pareto optimal solution
- If the problem is convex, all Pareto solutions can be found by selecting different weights
- Some methods (Weighted Chebyshev, Exponential Weighted Criterion) may be able to find all Pareto solutions for non-convex problem, but may also produce non-Pareto optimal solutions

It is very important to check the convex conditions if you want to produce the Pareto set using weighting methods.

3.2 $\epsilon$-constraint method [9]

The idea of $\epsilon$-constraint method is very straight forward: choose one objective to optimize and treat other objectives as constraints with pre-determined upper bounds:

$$\min_{\mathbf{x}} f_l(x) \tag{10}$$
$$\text{subject to} \quad f_i(x) \leq \epsilon_i, \text{ for all } i \neq l,$$
$$\epsilon_i \text{ is a known the upper bound of } f_i$$

Here $f_l$ is the one objective to optimize. If the solution of (10) is unique, the solution is Pareto optimal for the original MOO problem (2) [1]. By setting different upper bounds $\epsilon_i$, this method may get a complete Pareto set under both convex and non-convex situations. But if $\epsilon_i$ is too small, (10) may not have any solution since there may be no feasible solutions.

3.3 Normal Boundary Intersection (NBI) [10] & Normal Constraint (NC) method [11]

Both NBI and NC methods follow the same approach to find Pareto set:

    Step 1: Find *Anchor Points* in objective space
    Step 2: Define *Utopia Line*(hyperplane) connecting Anchor Points
    Step 3: Set evenly distributed *base points* on Utopia Line
    Step 4: Optimize one objective in area above *normal vectors* at each base point

An anchor point $A_i$ is a point in objective space with one objective to be minimized:

$$A_i = F(x^{*i})$$

where solution $x^{*i}$ minimize $i$th objective $f_i$: $x^{*i} = \arg_x \min f_i(x)$. We use the Figure 8 to illustrate these steps:

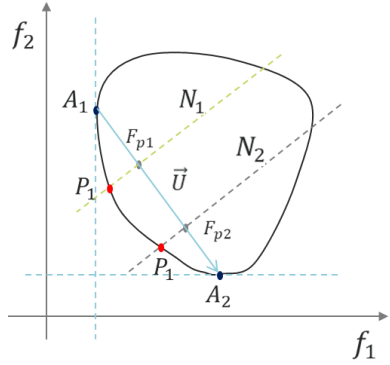

Figure 8. NBI & NC Methods

In Figure 8, $A_1$ and $A_2$ are two anchor points. The Utopia line vector $\vec{U}$ is the line connecting $A_1$ and $A_2$:

$$\vec{U} = A_1 - A_2$$

Two base points $F_{p1}$, $F_{p2}$ on Utopia line are defined by:

$$F_{pj} = \omega_{1j} A_1 + \omega_{2j} A_2$$

Let $\vec{N_1}$ and $\vec{N_2}$ be two normal vectors at two base points $F_{p1}$, $F_{p2}$. The following constraint defines a feasible region above the normal vector:

$$E_i = \{x \,|\, \vec{U} \cdot (F(x) - F_{pj}) \leq 0\}$$

The single objective optimization problem is defined with this constraint:

$$\min_{x \in E_i} f_2(x) \tag{11}$$

This method can find all the Pareto optimal solutions by specifying evenly distributed based points on the Utopia line, no matter the original MOO problem is convex or not. However, it also generates non-Pareto optimal solutions if the Pareto front is concave. A special filtering process has been employed in NC method to filter out non-Pareto solutions.

3.4 Other scalarization methods

Other scalarization methods can be found in the MOO literatures. Table 1 is a summary of three methods. Interested readers can get more details from related references:

| Method | Idea | Scalarization | Characteristic |
|---|---|---|---|
| Goal Programming [12] | Set up goal for each objective | $\min \sum_{i=1}^{M} |d_i|$ | May not be Pareto optimal |
| Physical Programming [13] | Map goals and objective to utility functions $\bar{g}_i$ | $\min \log_{10} \sum \bar{g}_i$ | Pareto optimal Need detail knowledge of each objective |
| Lexicographic Method [14] | Order each objective by importance | Minimize each objective in order | The solution may not be feasible |

Table 1. Other Scalarization Methods

## 4. Multi Objective Evolutionary Algorithms (MOEA)

MOEA is a class of optimization algorithms that emulate nature evolution processes, such as gene evolution, a flock of birds finding essential resources, a natural melting crystal cooling process and a family of ants seeking the best path for foods. These natural evolutions optimize themselves during the process using various techniques, which inspire different evolutionary algorithms. The evolutionary algorithms started in early 1950's and became a mature research field at the beginning of 1990's. Before 1985, when David Schaffer published the first paper of solving MOO problem using a genetic algorithm, the evolutionary algorithms research focused on solving single objective optimization problem. In 1989, based Schaffer's work, Goldberg introduced two principles to solve MOO problem using evolutionary algorithms: using dominated relation to assign fitness solution for each solution, and fitness value need to be adjusted to maintain the diversity of the final solutions. Many popular MOO EA methods were developed following these two principles. In this section, we will first explain the basic concepts of evolutionary algorithms. Then we will discuss more details on different MOO generic algorithms.

4.1 Evolutionary Algorithm (EA) Basic

To emulate a nature evolutionary process, EA uses the following terms to represent an optimization problem:

- *Individual*: a feasible solution $x$
- *Population*: a set of feasible solutions
- *Parents*: a sub-set selected from Population
- *Children*: new solutions produced from Parents

There are four operators applied to each set of solutions:

- *Evaluation*: assign fitness value $z(x)$ to each individual in Population
- *Selection*: select Parents from population, usually a random process based on fitness value, or a tournament process based on dominance relations in MOO
- *Variation*: mutation or crossover from Parents to produce Children
- *Elitism*: maintain 'better' solutions in next generation Population

These different set of solutions and their operators can be illustrated by in Figure 9:

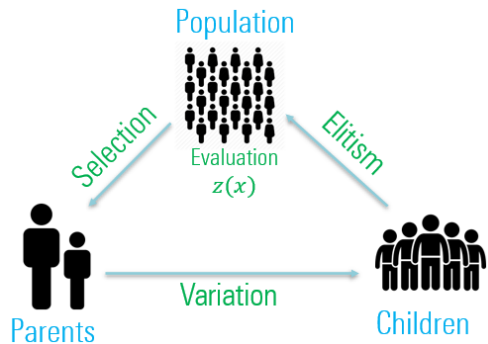

Figure 9. Evolutionary Algorithms

Figure 10 is a genetic algorithm example that shows the iteration workflow [15]:

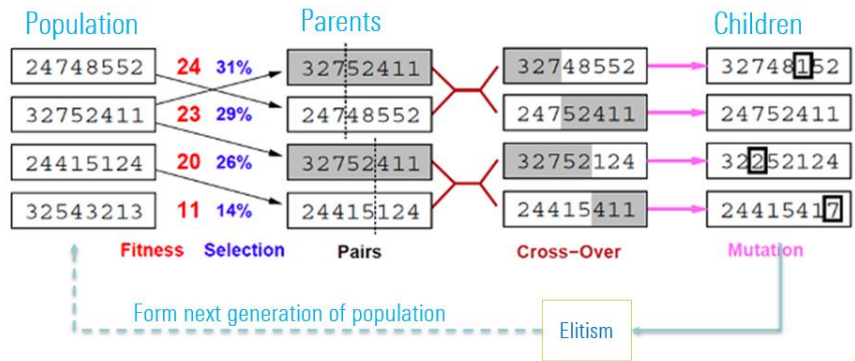

Figure 10. Genetic Algorithm Workflow

Before using EA to solve a real-world problem, we need to map a real-world solution to an individual. This is called *encoding*. Here we explain three methods to encode a solution in recommender system for Genetic Algorithm (GA):

1) Binary Encoding
2) Permutation Encoding

3) Real Value Encoding

In GA, an individual is a chromosome representing a solution, which is a recommendation list in recommender systems. In *binary encoding*, the value of each gene in the chromosome is 0 or 1. The length of the chromosome is number of available items. The sum of all the gene value should be number of items in the recommendation list. Binary encoding can be shown in Figure 11:

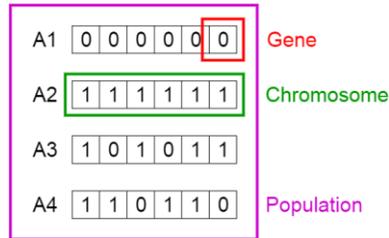
Figure 11. Binary Encoding

If the number of available items is too big, such as in the online sale platform, the chromosome becomes too long to handle. To solve this problem, *permutation encoding* uses the index of the item to fill the value of a gene, while the length of a chromosome is the length of recommendation list, as shown in Figure 12.

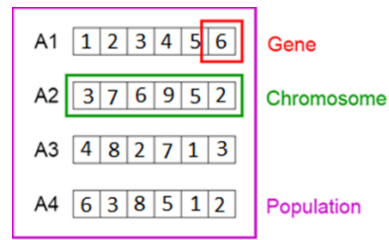
Figure 12. Permutation Encoding

The last encoding method is *real value encoding,* which is used to optimize the set of real value parameters. Each gene has the value of the parameter as shown in Figure 13.

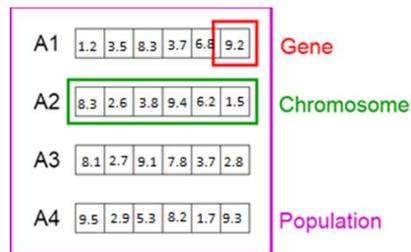
Figure 13. Real Value Encoding

Real value encoding is often used to find the best weights in the scalarization algorithms.

4.2 From single objective EA to multi objective EA

EA was first developed to solve a single objective problem. To solve a MOO problem, EA must find a way to determine which solution is 'better' given multiple objectives. All operators, except elitism, has to deal with the order or 'goodness' of the solution. As we discussed at the last section, MOO uses the dominance relation to decide the order or goodness of each solution. Under this consideration, at least one of three operators, evaluation, selection, and elitism, must use dominance relation to solve an MOO problem. Table 2 is a summary of major MOO GAs and where dominance relation is applied.

| Method | Fitness Evaluation | Parent Selection | Elitism |
| --- | --- | --- | --- |
| VEGA [16] | Single objective | Probability distribution | Dominance relation |
| MOGA [17] | Dominance relation | Probability distribution | NA |
| NPGA [20] | No Fitness | Tournament method (dominance relation) | NA |
| PAES [21] | No Fitness | Local search (dominance relation) | Dominance relation |
| NSGA [18] | Dominance relation | Probability distribution | NA |
| NSGA-II [19] | Dominance relation | Probability distribution | Dominance relation |

Table 2. MOO Algorithms Based on Genetic Algorithm

In the following sections, we will focus on the operators where dominance relations are used.

4.3 Vector Evaluated Genetic Algorithm (VEGA) [16]

VEGA is the first evolutionary algorithm used to solve a MOO problem. It first randomly generates a set of feasible solutions as the initial population. Then it randomly partitions the whole population into $M$ number of sub population $P_i$, where $M$ is the number of objectives. The fitness value of solutions is each $P_i$ is assigned by the $i$th objective value:

$$z_i(x) = f_i(x) \quad \text{for } x \in P_i$$

With fitness value, the parents are selected based on the probability:

$$p_i(x) = 1 - \frac{z_i(x)}{\sum_{s=1}^{M_i} z_i(x_s)} \qquad (12)$$

where $M_i$ is the number of solutions in sub population $P_i$. A solution with smaller fitness value has a high probability to be selected as parent. After regular mutation, the non-dominated children and parents will be selected to form the next generation. VEGA is simple but suffers the issue of concentrating to some extreme solutions.

4.4 Multi-Objective Genetic Algorithm (MOGA) [17]

In MOGA, each solution $x$ in the Population is ranked based on the number of solutions that dominates $x$: $r(x) = 1 + $ (# of solutions that dominates $x$). Then a fitness value is defined as

$$z(x) = N - \sum_{k=1}^{r(x)-1} n_k - 0.5(n_{r(x)} - 1) \qquad (13)$$

where $N$ is the number of solutions in population and $n_k$ =Number of solutions with $r(x) = k$. If $x$ is a non-dominated solution, then $r(x) = 1$, and $z(x) = N - 0.5(n_1 - 1)$. So all the non-dominated solution has the same largest fitness value. The effectiveness of (413) can be illustrated by Figure 14:

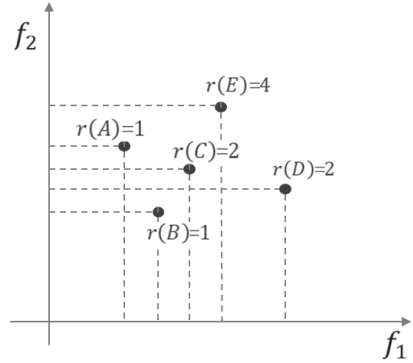

Figure 14. Ranking in MOGA

In Figure 14, $N = 5$. The rank $r(x)$ is labeled for each solution. Then we have:

$$n_1 = 2, \quad n_2 = 2, \; n_3 = 0, \; n_4 = 1$$

And

$$z(E) = 5 - (2 + 2) - 0.5(1 - 1) = 1$$
$$z(C) = z(D) = 5 - (1) - 0.5(2 - 1) = 3.5$$
$$z(A) = z(B) = 5 - 0 - 0.5(2 - 1) = 4.5$$

Both non-dominated solution $A$ and $B$ has the same maximum fitness value. $E$ has the smallest fitness value, representing the 'worst' solution.

4.5 Nondominated Sorting Genetic Algorithm (NSGA) [18]

NSGA sorts all solutions in Population by selecting non-dominated solutions in a iterative way. After moving the non-dominated solutions from initial population into to sub population, it gets the next non-dominated solution set from the remaining population. After this process ends without any remaining solution left, it gets a set of ordered sub-population set, with solutions in the previous sub-population dominates those in the next sub-population. Here is an example to show how this sorting process work in Figure 15:

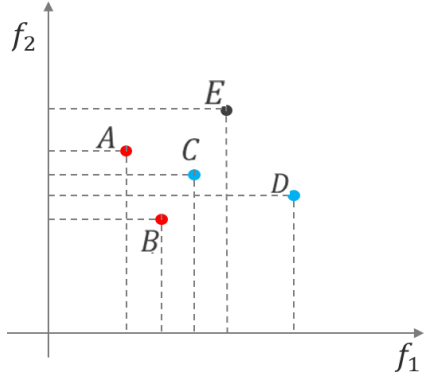

Figure 15. Nondominated Sorting in NSGA

In Figure 15, the initial population: $P = \{A, B, C, D, E\}$. We have

First set of non-dominated solutions: $P_0 = \{A, B\}$
Second set of non-dominated solutions: $P_1 = \{C, D\}$
Third set of non-dominated solutions: $P_2 = \{E\}$

Then we assign a dummy fitness value for solutions in each $P_i$ with descending order:

$$P_0: z(A) = z(B) = 10$$
$$P_1: z(C) = z(D) = 8$$
$$P_2: z(E) = 5$$

These fitness value can reflect how 'good' each solution based on the dominance relationship. A big advantage of NSCG method is that all the dominated solutions have the same fitness value. This will help to find all Pareto optimal solutions.

4.6 Nondominated Sorting Genetic Algorithm II (NSGA-II) [19]

NSGA-II in an improvement of original NSGA method in both sorting and diversity maintained. The so called 'fast non-dominated sorting' method use the number of dominated solutions for each solution $x$ to partition the population. The first sub-population $P_1$, contains solutions that has zero number of dominated solutions. This is equivalent to all the non-dominated solutions. The second sub-population $P_2$ contains solutions that has 1 dominated solution. The rest sub-population $P_i$ will be defined in similar way. The sub-population set $P_1, P_2, \ldots, P_N$ form an ordered set based on dominance relationship.

For the example in Figure 15, we have partition:

$$P_1 = \{A, B\}, \quad \text{0 dominated solution}$$
$$P_2 = \{C, D\}, \quad \text{1 dominated solution}$$
$$P_3 = \{\}, \quad \text{2 dominated solutions}$$
$$P_4 = \{E\}, \quad \text{3 dominated solutions}$$

Same as in NSGA method, we assign fitness for solutions in each sub-population based on their order:

$$P_1: z(A) = z(B) = 1$$
$$P_2: z(C) = z(D) = 2$$
$$P_4: z(E) = 3$$

Because of its sorting efficiency NSGA-II becomes a state-of-art MOEA method to solve MOO problem.

4.7 Niched Pareto Genetic Algorithm (NPGA) [20]

There is no fitness value assignment in NPGA. It uses a tournament method to select parent instead. We use the example of Figure 15 to explain the idea. In Figure 15, there are five solutions in the population $P = \{A, B, C, D, E\}$. First, we randomly select a comparison group: $G = \{C\}$. For $\forall\, x, y \in P$, choose a 'better one' comparing with $G$:

- $(A, E)$:   $E$ is dominated by $G$, choose $A$
- $(B, E)$:   $E$ is dominated by $G$, choose $B$
- $(B, D)$:   neither $B$ or $D$ is dominated by $G$. choose $D$ in the less crowded region.

The parent will be selected as $A, B, D$. Since the tournament method can use dominance relationship and maintain the diversity at the same time, it can produce more evenly distributed solution set.

4.8 Pareto Archived Evolution Strategy (PAES) [21]

PAES is another MOGA that does not assign fitness value to solutions. PAES uses a dominance relation based local search and a tournament method to find better solutions. We explain this idea using the example in Figure 16.

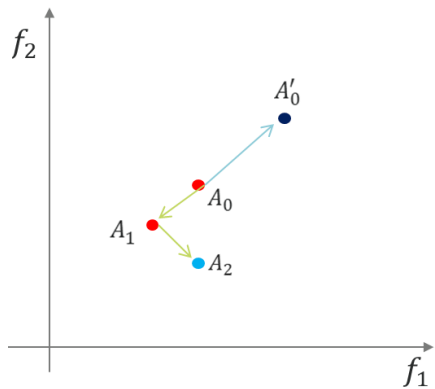

Figure 16. Pareto Archived Evolution Strategy

We first set aside an empty set $G$ to hold all the non-dominated solutions found later. We choose a solution $A$ as the parent and add it to $G$: $G = \{A_0\}$. The we start the local search:

Mutate $A_0$, get child $A'_0$, which is dominated by parent $A_0$, then discard $A'_0$.
Mutate $A_0$, get child $A_1$, which dominates parent $A_0$, $A_1$ become the new parent

Mutate $A_1$, get child $A_2$, no dominance relationship between $A_1$ and $A_2$ , <u>equally good</u>

When child and parent are equally good, we need to update archived solution set $G$:

- Replace solutions with $A_2$ in G if they are dominated by $A_2$
- Chose $A_2$ or $A_2$ from the less crowded region if no solution in $G$ is dominated by $A_2$

In this example, $A_0$ in $G$ is dominated by $A_2$. So $A_0$ is replaced by $A_2$. With newly updated $G = \{A_2\}$, starting a new iteration. The solution is $G$ will be the final solution set.

4.9 Maintaining solution diversity

All MOO evolutionary methods inherit a major issue from original evolutionary algorithms: converge to local optima rather than a global optimum. In MOO, such drawback may generate un-evenly distributed Pareto set. How to maintain the diversity of the final solution set is the problem every evolutionary algorithm needs to solve. As we discussed in previous sections, the niche count and fitness sharing method [22] is commonly used in all MOEAs:

- Niche count $nc(x)$: a measurement of crowdedness around the solution $x$
- Fitness Sharing $z'(x)$: discount fitness value with niche count

$$z'(x) = \frac{z(x)}{nc(x)} \qquad (14)$$

With fitness sharing, solutions in more crowded region are less likely to be selected. Thus maintain the diversity of the final solution set. Interested readers may find more details in the related references.

4.10     MOEA methods based on other EAs

Except genetic algorithms, other evolutionary algorithms can also be used to solve MOO problems. Here is list of these MOEA methods.

- PSO based MOEA [23]
- SA based MOEA [24]
- ACO based MOEA [25]

We will not get details of each of these MOEAs in this paper.  Interested readers can get more knowledge about these methods by checking related references.

# 5. Selection of Best Solutions in Pareto Set

In both scalarization algorithms and evolutionary algorithms, if no DM preferences for objectives are available, these algorithms need to produce the Pareto set. In a posteriori cases, DM can select a best solution from Pareto set. However, if no DM is available, or DM is not capable to select from Pareto set, we must select a 'best' solution for DM. This is most cases in recommender systems.

Theoretically, we can only 'guess' be the best solution for DM if DM is not available. There are several categories of these guessing methods:

- Knee point methods
- Hypervolume methods
- Multiple-criteria decision-making (MCDM) methods

We will highlight some of these methods in the following sections.

5.1 Knee point methods

A knee point is special point in Pareto front at position like a 'knee'. A general description of a knee point is "*a small improvement in either objective will cause a large deterioration in the other objective*". This means, without other information, a DM does not want to move away from kneed point since he/she may lose more in other objectives if take a small gain in one objective. A geometric illustration in a 2-objective case is shown in Figure 17.

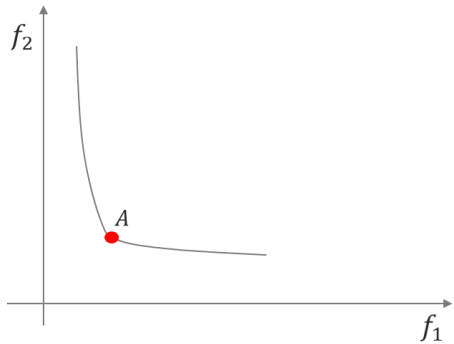
Figure 17. Knee Point

There are many methods to determine or find a knee point given a set of solutions in Pareto front. The simplest method is called *Angle Based Knee Point* [26], which can only be used in two-objective problems.

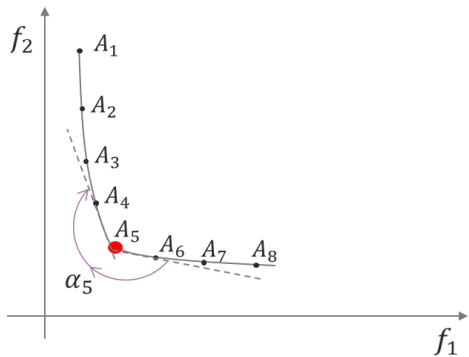
Figure 18. Angle Based Knee Point

In Figure 18, Pareto front is given by $P = \{A_1, A_2, \ldots, A_8\}$. We calculate reflex angle $\alpha$ for each point in $P$: $\alpha_i = \angle A_{i-1}A_iA_{i+1}$. The solution with maximum reflex angel is the knee point:

$$\alpha_5 = \max\{\alpha_1, \alpha_2, \alpha_3, \alpha_4, \alpha_5, \alpha_6, \alpha_7\}$$

So $A_5$ is the knee point in this example.

Other knee point methods, such as *Marginal Utility Method* [26], *Hyperplane Normal Vector Method* [27] can be found in the related references.

5.2  Hypervolume methods [28]

For a given solution point $F(x)$ in Pareto front in two objectives, the hypervolume of solution $x$, $V(x)$, is the area of rectangle of origin to $F(x)$ as shown in Figure 19.

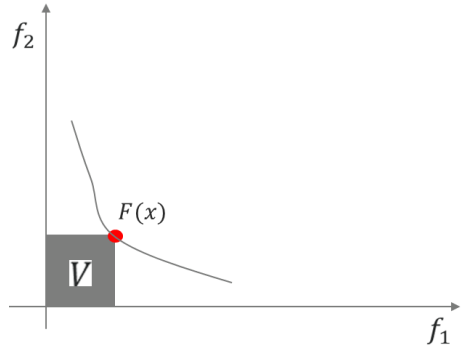

Figure 19. Hypervolume Method

If the Pareto front is convex, the solution with the maximum hypervolume will is the best solution:

$$\max_x V(x) \qquad (15)$$

The hypervolume methods can also be used to evaluate the effectiveness of Pareto set by calculating the hypervolume of the Pareto set:

$$V(P) = \sum_{x \in P} area\ of \bigcup V(x) \qquad (16)$$

If there are two Pareto set $P_1$ and $P_1$ for the same MOO problem, the Pareto set with a smaller hypervolume $V(P_1)$, $V(P_2)$ is better.

5.3  Multiple-criteria decision-making (MCDM) methods

Multiple-criteria decision-making (MCDM) or Multiple-criteria decision analysis (MCDA) is a dedicated discipline that try to evaluate conflicting criteria in decision making process. Given a set of Pareto solutions, how to select the best solution is the same decision-making process. The only difference is that most MCDM methods require DM preference input, while DM is not available in our situation. We can set the equal preference to all objectives in MCDM methods where is needed. In the following we just highlight two MCDM methods: *Technique for Order of Preference by Similarity to Ideal Solution* (TOPSIS) [29] and *Preference Ranking Organization METHod for Enrichment of Evaluations* (PROMETHEE) [30].

The idea behind TOPSIS method tries to find solution that has minimum distance to the ideal Utopia point and has maximum distance to the worst up bound point:

$$\min_x ||F(x) - F^*|| \qquad (17)$$
$$\max_x ||F(x) - F^-|| \qquad (18)$$

where $F^*$ is the Utopia point, which takes the minimum of each objective. $F^-$ is the up-bound point, which take the upper bound of each objective.

PROMETHEE method uses a pairwise ranking method to evaluate the objective difference of pair of solutions. After normalizing these differences into value in [0, 1], taking the weighted sum of normalized difference from each objective, $\pi(a, x)$, both positive $\phi^+(a)$ and negative $\phi^-(a)$ preference scores are calculated for each solution $a$:

$$\phi^+(a) = \frac{1}{|P| - 1} \sum_{x \in P} \pi(a, x) \qquad (20)$$

$$\phi^-(a) = \frac{1}{|P| - 1} \sum_{x \in P} \pi(x, a) \qquad (21)$$

These scores represent how solution $a$ is preferred comparing to other solutions from positive and negative directions. The final net preference score is calculated by:

$$\phi(a) = \phi^+(a) - \phi^-(a) \qquad (22)$$

which represents the overall favorite of solution $a$. A solution with higher net preference score represents a better solution.

Figure 20 shows how $\pi(a, x)$ is calculated:

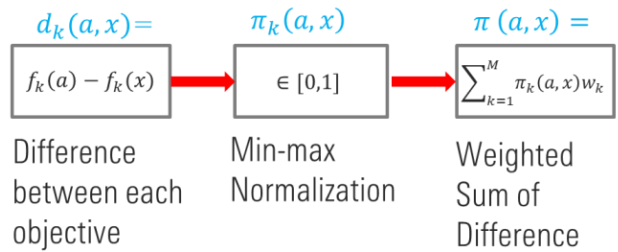

Figure 20. Pairwise Ranking Calculations

## 6. MOO Libraries

Developing MOO algorithms from scratch is always a dauting and time-consuming task. Fortunately, there are several available MOO libraries. We can always choose the one that best fit your needs.

Table 3 shows commonly used MOO libraries and their information:

| Name | MOO Methods | Single Objective | Language | Open Source | Last update |
|---|---|---|---|---|---|
| PyGMO[2] | NSGA-II, MOEA, MH-AOC, NS-PSO | Yes | Python | Yes | 2021 |
| pymoo[3] | NSGA-II, NSGA-III, MOEA | Yes | Python | Yes | 2020 |
| Inspyred[4] | PAES, NSGA-II | Yes | Python | Yes | 2019 |
| Platypus[5] | NSGA-II, MOEA, SPEA2, MOEA/D, PSO, PAES, PESA2 | No | Python | Yes | 2019 |
| MOEA Framework[6] | NSGA-II, NSGA-III, PAES, PESA2, SPEA2, MOEA, MO-PSO | Yes | Java | Yes | 2019 |
| MATLAB & Simulink[7] | MOEA, NSGA, SPEA2 | Yes | Matlab | No | 2021 |
| openGA[8] | NSGA-III | Yes | C++ | Yes | 2020 |

Table 3. MOO Libraries

Most of these MOO libraries are open source and free to use. Please note, there is no available libraries for scalarization algorithms since these they can be solved by a single objective optimizer. The 'Single Objective' in the above table show if the library includes single objective optimizers.

---

[2] https://esa.github.io/pygmo2/index.html
[3] https://pymoo.org/
[4] https://pythonhosted.org/inspyred/
[5] https://platypus.readthedocs.io/
[6] http://moeaframework.org/
[7] https://www.mathworks.com/matlabcentral/fileexchange
[8] https://github.com/Arash-codedev/openGA

## Summary

In this section, we introduced the MOO basic concepts and algorithms to solve MOO problems. The special challenge of MOO problem is how to rank the solutions with multi objectives. The dominance relation establishes a semi order for all feasible solutions of MOO problems. With dominance relation, the non-dominated solutions, also known as Pareto optimal solutions, should be selected as the optimal solutions.

If DM preferences about each objective can be quantified as weight factors, then a Pareto optimal solution can be found by solving the scalarization problem transformed by defined weights, under the condition that the sum of weight equals to one and each weight is positive. This is known as a priori. If DM cannot set preferences for each objective, then the MOO algorithms must find all Pareto optimal solutions, which is Pareto set. Both scalarization algorithms and MOEA can be used to generate Pareto set. However, you need to run scalarization algorithms multiple times with different parameters to get the Pareto set. In many cases, convex condition is necessary to get the complete Pareto set. Population based MOEA can produce Pareto set in one run. The solution diversity needs to be maintained in order to get evenly distributed Pareto set.

For many MOO problems, such as MOO recommender systems, DM is not able to choose the best solution from Pareto set. Then we need to 'guess' and find a best solution for DM. This is especially important in recommender systems with multi objectives.

# III. Recommender Systems with Multi-Objective Optimization

In this section, we discuss multi-objective recommendations.

## 1. Recommender Systems: Single-Objective vs. Multi-Objectives

Several recommendation algorithms have been proposed to produce personalized recommendations. One of these algorithms is the model-based approaches in which we need to utilize the optimization techniques in machine learning. In traditional recommender systems, we usually optimize a single objective, e.g., minimizing the prediction errors in the rating prediction task or maximizing the ranking quality in the top-N recommendation task.

Using MOO in recommender systems is not a novel practice. The earliest application of MOO in recommender systems may track back to the ones which balance different evaluation metrics (e.g., accuracy, diversity, novelty, etc.). Recently, there is an emerging demand in MOO, especially in some special type of recommender systems. Take the multi-task recommender system [31] for example, researchers may utilize a joint learning to optimize multiple tasks in a model with shared representations (e.g., latent factors, feature embeddings, etc.).

We have identified six contexts in which we may need MOO in the area of recommender systems, as shown by the Figure 21. We discuss them in details in the next section.

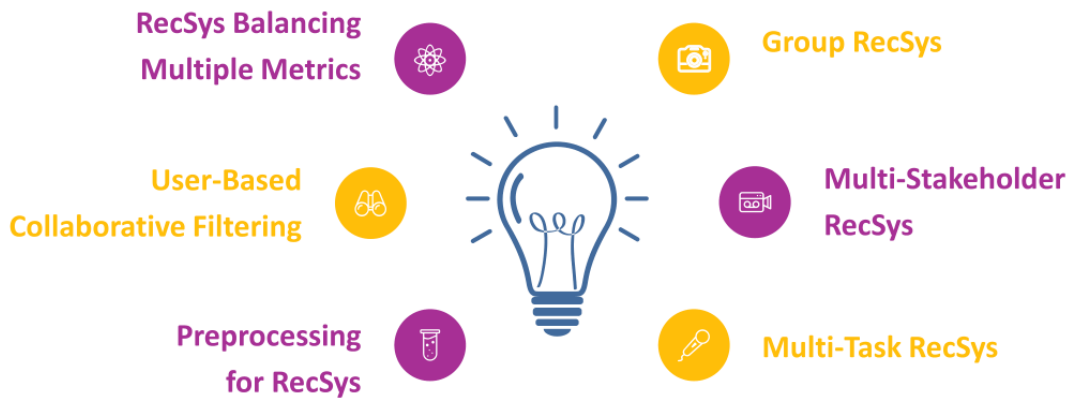

Figure 21 Six categories of Multi-Objective Recommendations

## 2. Multi-Objective Recommendations

In this section, we will discuss multi-objective recommendations according to the six categories above. For each category, we cover the motivation (e.g., why we need MOO in this category), objective definitions (how to generally define objectives in this category), goals and case studies.

**2.1 Recommender Systems Balancing Multiple Metrics**

Accuracy or relevance may not be the only focus in recommender systems. Take news or music recommendations for example, users may be boring if the system always recommends the same types of the news or music. Users may would like to try something different (i.e., diversity) or something new (i.e., novelty). Kaminskas, et al. [32] discussed different types of the evaluation metrics which may be considered in the recommender systems, including diversity, novelty, serendipity, coverage, and so forth. The recommendation problem, in this case, results in a multi-objective problem in which the goal is balancing multiple metrics, or improving other metrics at no or acceptable loss on accuracy of the recommendations. Note that the definition of these metrics, such as novelty and diversity, may vary from applications to applications. For more details about the definitions or equations related to these metrics, refer to Kaminskas, et al.'s work [32].

In this category, most of the research utilized MOEA as the MOO to solve the problems. Take Chai, et al. [33]'s work for example, they built item-based collaborative filtering to produce the top-K recommendation list for each user first. Afterwards, they utilized a multi-objective immune algorithm (MOIA) to learn a top-N recommendation list for each user, while the top-N items will be learned and selected from the top-K items (N<<K). The MOIA can learn such top-N recommendation lists by considering accuracy and diversity as the multiple objectives, and finally produce a Pareto set. The authors tried to use PROMETHEE [30] to select the single best solution from the Pareto set. It is worth mentioning that the framework in Chai, et al. [33]'s work is general and typical, while most of the research balancing multiple recommendation metrics utilized a similar process (i.e., learning recommendation list directly through MOEA). Note that the process of producing top-K recommended items is optional. It is used to reduce the computation cost in MOIA by refining the list of recommendation candidates.

The work by Ribeiro, et al. [34] provided another way to utilize MOEA in recommender systems by considering accuracy, novelty and diversity. More specifically, they built multiple rating prediction recommendation models, and aggregated the predicted ratings by a linear weighted sum. The aggregated rating can be used to produce the top-N recommendation list so that they can measure accuracy, diversity and novelty. They utilized MOEA to learn the weights in the hybrid model (i.e., linear aggregation) by considering the dominance relations among accuracy, diversity and novelty. They finally defined different importance of the objectives, and used a weighted sum (i.e., importance as the weights) of the objective values to select the optimal solution from the Pareto set. In contrast to other research work in this category, they did not learn the recommendation list directly, but utilized MOEA to learn the weights in the hybrid recommendation model. However, they manually tried different sets of the importance associated with the objectives to select the single optimal solution. More formal approaches for the selections can be found in Section II.

### 2.2 User-Based Collaborative Filtering

The goal of the research by Karabadji, et al. [35] is to improve accuracy and diversity in the recommendations. By contrast, they proposed an approach which was specifically designed for user-based collaborative filtering (UBCF). The underlying assumption in their work is that the diversity of the recommendations in UBCF is dependent with the diversity of the user neighborhood. As a result, they proposed their approach in which they select the user neighborhood by considering two objectives -- the similarity between the target user and the user neighbors, and the dissimilarity among the user neighbors (i.e., the intra-group diversity in the neighborhood). They utilized a weighted sum as the scalarization method to aggregated these two objectives, and adopted genetic algorithm to learn the user neighborhood, as the MOO process. UBCF will take advantage of the learned neighborhood to produce recommendations. Karabadji, et al. tuned up the parameter (i.e., the weights in the scalarization) by evaluating the recommendation performance.

### 2.3 Preprocessing for Recommender Systems

Some recommendation models may utilize the outputs from an unsupervised learning process. For example, a clustering process may be applied to produce user or item clusters, while recommender systems can be built upon these clusters. As a result, the MOO process could be applied to these preprocessing stages. Usually, it is involved with a process of clustering or association rules. MOO can be used to produce better outputs which can further assist recommender systems.

In the partitional clustering, the inter-cluster and intra-cluster distances can be considered as two objectives to be optimized. In the process of rule mining, support and confidence could be two objectives to be considered.

Take Neysiani, et al.'s work [36] for example, they considered support and confidence as two objectives to produce association rules like $(T_1, T_2) \rightarrow T_3$. The rules can be further used to produce recommendations, e.g., if a user likes T1 and T2, the rule above may infer that the user may also like the item $T_3$. They utilized the weighted sum as the scalarization method, and adopted genetic algorithm to learn the association rules which will be further applied to generate top-N recommendations.

## 2.4 Group Recommender Systems

In contrast to traditional recommender systems, the group recommender systems [37] produce a list of recommended items to a group of users, e.g., group dining or travelling. The major challenge in group recommendations is to balance the individual and group satisfactions, since a member's taste in the group may conflict with another member's preferences in the same group. As a result, MOO can be applied to find a balance and improve group recommendations.

The work by Xiao, et al. [38] utilized the weighted sum as the scalarization method to combine two objectives – individual satisfaction and group fairness. They proposed to use greedy search and integer programming as the single objective optimizer. Their experiments were able to demonstrate that they could improve group recommendations by considering group fairness in the MOO process.

## 2.5 Multi-Stakeholder Recommender Systems

Multi-stakeholder recommender system [39] is another type of the recommender systems in which researchers try to balance the needs of multiple stakeholders. Traditional recommender systems may produce the recommendations by considering the preferences of the end users only. However, the receiver of the recommendations may not be the only stakeholders. For example, buyers and sellers are two stakeholders on eBay.com, while students, instructors, parents and publisher may be the possible stakeholders in the context of book recommendations in educations. The recommendations should be produced and delivered by considering the utility of the items from the perspective of different stakeholders. In this case, each stakeholder may be involved with at least one benefit or objective. MOO becomes the general solution to balance the needs of multiple stakeholders.

The definition of the objectives in this category may vary from applications to applications. In the marketplace, buyers, sellers and the platform may be the stakeholders. In job seeking, the recruiter and the job seekers may be the stakeholders. Therefore, the definition of the stakeholders and the associated objectives are dependent with the specific applications or domains.

Lin, et al. utilized scalarization to optimize the click through rate (CTR) and gross merchandise volume (GMV) in an e-commerce application [40]. They defined the loss function based on CTR and GMV, and used the weighted sum to combine these two losses into a joint loss function. They tried different weights to perform the joint learning process and finally adopted the least misery strategy to select a single optimal solution from the Pareto set.

By contrast, Zheng, et al. utilized MOEA as the optimizer to recommend Kaggle datasets to students for the projects by considering the item utility from the perspective of both students and instructors [41]. They took advantage of the multi-criteria rating vectors and expectation vectors to compute the item utility in view of students and instructors by using a utility-based multi-criteria recommendation framework [42]. Weighed sum is used to aggregate the two utilities (i.e., from perspective of students and instructors) to produce the ranking score which can be applied to sort and rank items. MOEA was applied to consider the student and instructor satisfactions, as well as the recommendation performance as the multiple metrics. A TOPSIS based method [29] was used to select a single optimal solution from the Pareto set generated by MOEA. The experimental results can demonstrate that the model could improve the instructor's satisfaction at a limited loss on the recommendation performance. A demo of this research by using the open-source MOEA framework[9] can be found here[10].

### 2.6 Multi-Task Recommender Systems

The multi-task recommender systems [31] refer to the recommender system which optimize multiple tasks through a joint learning process. Particularly, there are usually some common or shared representations in multi-task recommender systems, such as the shared latent-factors or embedding layers in the neural network models. Note that, multi-task recommender systems are not limited to the models using neural networks. Some work utilizing matrix factorization can optimize both rating prediction and ranking tasks by sharing the user and item latent factors. The goal is to improve multiple tasks by a joint learning process. However, the improvement is dependent with the correlation of the tasks and the power of the shared representations.

Each task may be associated with at least one objective. Therefore, the joint learning process in the multi-task recommender systems can be considered as a process of MOO too. The definition of the multiple tasks may vary from applications to applications. Usually, there are three ways to build these joint tasks – first of all, it could be multiple recommendation tasks, such as the joint learning of a rating prediction task and a ranking task. In addition, the model may optimize multiple user feedbacks or behaviors, such as click-through, view-through, the number of purchases, impressions, and so forth. Furthermore, it is possible to combine a recommendation task with a non-recommendation task. For example, some research may produce recommendations together with a process of text productions, such as user reviews, opinions or the explanation of the recommendations. Or, it is possible to additionally predict trust or social relations, or run recommendation tasks along with a process of classifications or regressions.

---

[9] http://moeaframework.org/
[10] https://github.com/irecsys/Tutorial_MSRS

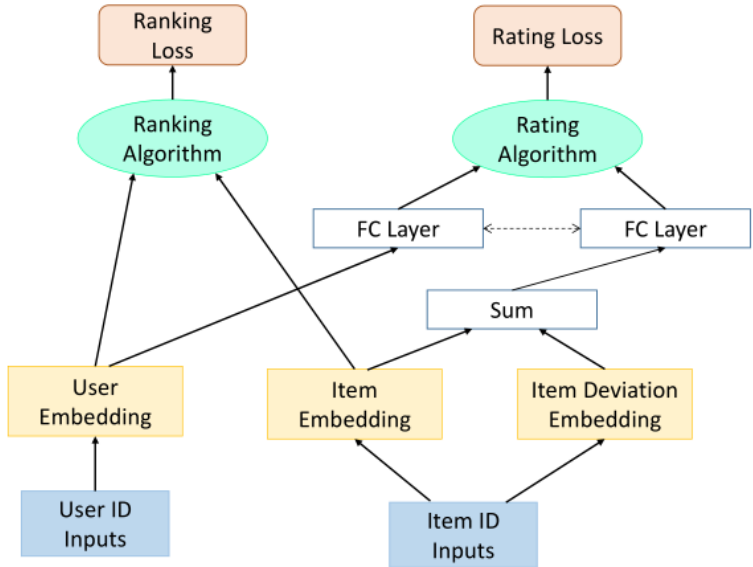

Figure 21 Joint Learning for Rating Predictions and Rankings [43]

The work by Hadash, et al. [43] shared the user and item embedding layers in the rating prediction task and ranking task. The joint learning is a process of scalarization by using the weighted sum to combine the loss function in the rating prediction and ranking tasks, as shown by Figure 21. Li, et al. [44] proposed a more complicated work which optimize the rating predictions and the tip generations, as shown by Figure 22. The user and item embeddings, the hidden layers in the process of review generations, as well as the predicted ratings, will be shared in the process of the tips generation which is realized by a gated recurrent neural networks (i.e., the right part in Figure 22). Weighted sum is used as the scalarization in the joint learning process by considering the rating prediction loss, review and tips generation losses.

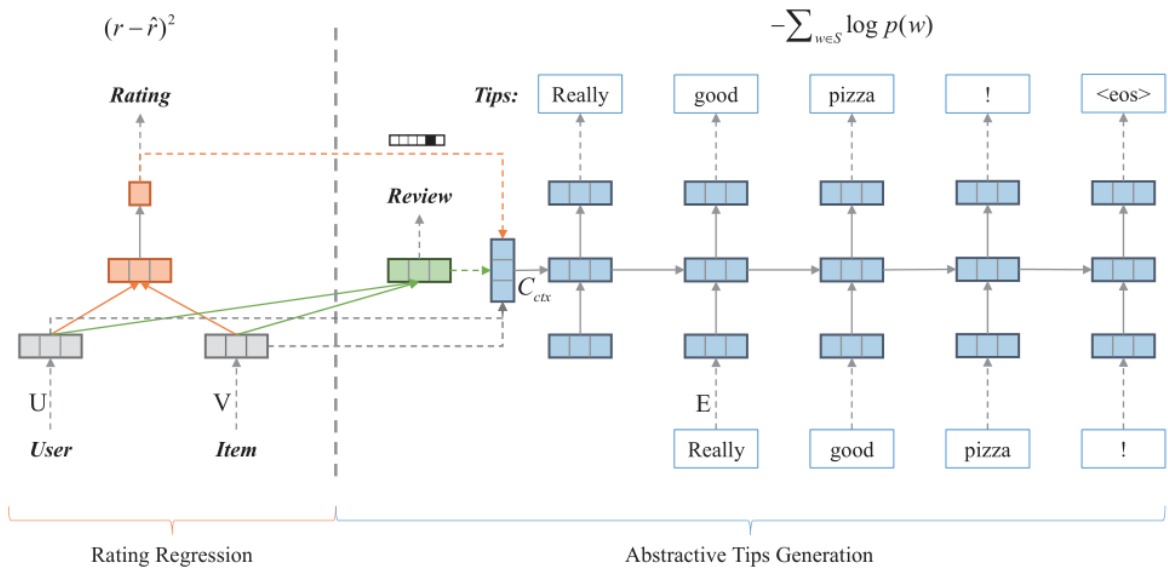

Figure 22 Joint Learning for Rating Regression and Tips Generation [44]

In multi-task recommender systems, most of the work utilized the joint learning process which is in shape of the weighed sum of the scalarization method. However, researchers may just assign a learning rate to each loss function without guaranteeing the sum of the weights is one. In this case, it is not guaranteed that researchers can find a Pareto set, though a single solution may still be Pareto optimal. Furthermore, the researchers may just tune up the learning and regularization rates to find a better solution than the baseline. According to the instructions in MOO, trying different weights to produce a Pareto set and selecting an optimal solution from the set should be formal way, if the preferences of the decision maker is not available.

## Summary and Challenges

The multi-objective recommendation is an application of MOO in the area of recommender systems, while we identify six contexts in which we may need MOO in recommender systems.

We provide a suggested workflow to utilize MOO in the area of recommender systems, as shown by Figure 23. Researchers should be clear about the multiple objective problems and define the multiple objectives in their applications. If the preferences of the decision maker are available, it can be considered as a single objective optimization process by using scalarization and any available single-objective optimizer. Otherwise, researchers can select scalarization or MOEA to produce a Pareto set, and try to find a single best solution from the Pareto set.

There are several challenges in multi-objective recommendations.
- *Which MOO method should be adopted? The scalarization or MOEA?*
  First of all, researchers should follow the suggested workflow shown by Figure 23. Scalarization is easy to be utilized in the process of joint learning. It may run fast and it is guaranteed to find a global optima due to mathematical foundations, especially when it comes to a convex problem. MOEA is easier to be applied or implemented, while there are several open-source libraries. However, the computation cost may be significantly increased if the data is large or the parameters to be learned are numerous. In addition, the process of heuristic search may converge to a local optima if the solution search is not well designed. However, MOEA is always a solution if the problem is non-convex.
- *How to select a single optimal solution from the Pareto set?*
  Different approaches to select a single optimal solution from the Pareto set has been discussed in Section II. However, there are limited research in recommender systems that compare different selection methods.
- *How to achieve a balance among the objectives?*
  In some research work, such as the recommender systems balancing multiple metrics or the multi-stakeholder recommendations, a balance is expected to be achieved among different objectives. Without explicit preferences of the decision makers, we suggest to utilize online user studies or surveys to evaluate the "balance".
- *How to present the effectiveness of the multi-objective models?*
  The goal of some research is to improve the recommendation quality by considering multiple objectives. Therefore, they only compare the models in terms of the recommendation performance without presenting or comparing multiple objectives in the experimental results. We do suggest researchers to compare different objectives in order to discover more insights about why the recommendations could be improved.

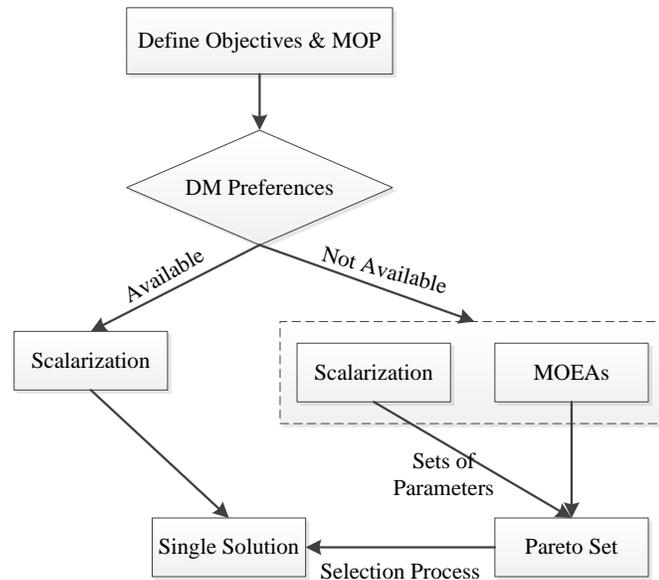

Figure 23 Suggested Workflow